# Quantum Conductors Formation and Resistive Switching Memory Effects in Zirconia Nanotubes

A.S. Vokhmintsev[1], I.A. Petrenyov[1], R.V. Kamalov[1], I.A. Weinstein[1,2]*

[1] NANOTECH Center, Ural Federal University, Ekaterinburg, 620002, Russia
[2] Institute of Metallurgy of the Ural Branch of the Russian Academy of Sciences, Ekaterinburg 620016, Russia

*e-mail: i.a.weinstein@urfu.ru

## Abstract

The development prospects of memristive elements for non-volatile memory with use of the metal-dielectric-metal sandwich structures with a thin oxide layer are due to the possibility of reliable forming the sustained functional states with quantized resistance. In the paper we study the properties of fabricated memristors based on the non-stoichiometric $ZrO_2$ nanotubes in different resistive switching modes. Anodic oxidation of the Zr foil has been used to synthesize a $ZrO_2$ layer of 1.7 μm thickness, consisting of an ordered array of vertically oriented nanotubes with outer diameter of 75 nm. $Zr/ZrO_2/Au$ sandwich structures have been fabricated by mask magnetron deposition. The effects of resistive switching in the $Zr/ZrO_2/Au$ memristors in unipolar and bipolar modes have been investigated. The resistance ratios $\geq 3 \cdot 10^4$ between high-resistance (*HRS*) and low-resistance (*LRS*) states have been evaluated. It has been founded the conductivity of *LRS* is quantized in a wide range with minimum value of $0.5G_0 \approx 38.74$ μ*S* due to the formation of quantum conductors based on oxygen vacancies ($V_O$). Resistive switching mechanisms of $Zr/ZrO_2/Au$ memristors with allowing for migration of $V_O$ in an applied electric field have been proposed. It has been shown that the ohmic type and space charge limited conductivities are realized in the *LRS* and *HRS*, correspondingly. We present the results which can be used for development of effective memristors based on functional $Zr/ZrO_2/Au$ nanolayered structure with multiple resistive states and high resistance ratio.

## Introduction

Today zirconium dioxide is being actively used in various high-tech fields such as the generation and storage of electrical energy [1, 2, 3], optical and laser technology [4], catalysis [5, 6, 7], photonics [8], solid-state dosimetry of ionizing radiation [9,10], nanoelectronics [11,12,13], biomedicine [3, 14, 15,], etc. The prospect of creating memristor elements of non-volatile memory based on metal-dielectric-metal (MDM) sandwich structures with a $ZrO_2$ layer is due to their outstanding properties. Among the latter are low energy consumption [16, 17, 18, 19, 20], high on/off current ratio [19, 20, 21, 22, 23, 24], high-speed performance [16, 18, 19, 20, 24, 25, 26, 27, 28] and meantime-to-failure (endurance), a variety of morphology and methods of manufacturing the active layer, the possibility of scaling [17, 18, 24, 28] and creating three-dimensional integrated circuits [21, 29], applications in flexible electronics [21], as well as compatibility with the existing CMOS technology [16, 30].

It is known that the resistive switching in the oxide layer of MDM structures underlying the memory device operation is usually provided by the mobility of anionic ($V_O$) vacancies [20, 25, 27, 28,29, 31, 32, 33], ions of impurity metals [18, 19, 20, 24, 26, 33, 34, 35] or $Zr^+$ [16, 28, 36] in the active layer under an external electric field. The electrical resistance of the memristor in low-resistance (*LRS*), high-resistance (*HRS*) and intermediate states are governed by the thickness and imperfection of the dioxide layer [23, 24, 28, 31, 32, 37]. In this case, one of the most probable mechanisms of resistive switching in as-grown $ZrO_2$-based structures is the chain-ordering of $V_O$ vacancies followed by the formation of conductive channels or filaments between metal contacts. Subsequently, in applying a control voltage $U$, the conductive filaments (CF) are partially destroyed and/or restored. Thus, a reversible switching of the memristor between the corresponding resistive states ($LRS \leftrightarrow HRS$) in both unipolar [22] and bipolar [36] modes is realized.

There are various methods for improving the characteristics of memristors based on functionalized $ZrO_2$ layers. To improve the stability and operational lifetime of memristor structures, in this context, the selection of materials for electrodes [21, 17, 25], the creation of nanocomposites with organic compounds [34], and combinations of several oxide layers [23] should be worth mentioning. Conductive metal filaments made of materials embedded in the active layer at the stage of electroforming the memristors can be provided by ion implantation [36, 26] and incorporation of metal nanocrystals into the oxide [24]. A rise in the degree of dioxide non-stoichiometry in oxygen makes it easier for further electroformation of Zr [16] or $V_O$ [22] filaments in the active layer.

An oxygen deficiency increase in various oxide structures during their synthesis can be achieved by the method of electrochemical oxidation through fitting conditions and anodizing parameters [21, 38, 39, 40, 41, 42]. In particular, it was shown in [38, 43, 44, 45, 46, 47, 48] that MDM structures based on nanotubular layers of non-stoichiometric $TiO_{2-X}$, obtained by the anodization technique, are breakthrough materials as memristor memory cells.

To date, only one scientific group has examined the effects of resistive switching in 40 nm thick $ZrO_2$ layers produced by the method of anodic oxidation at room temperature in the galvanostatic mode in an aqueous solution of phosphoric acid [21, 49]. Earlier, we have reported the results of studies of the current-voltage characteristics of nanotubular $ZrO_2$ layer-based as-grown MDM structures with unidirectional conductivity, obtained by the anodic oxidation method [37, 32] in the potentiostatic mode. The present paper is aimed at measuring and analyzing the static current-voltage ($I$-$V$) characteristics of fabricated nanoelectronic devices based on a $ZrO_2$ nanotubular layer in unipolar and bipolar resistive switching modes, and also at describing the peculiarities of electroformation of memristors and the latter's switching mechanisms.

## Experimental

### *Synthesis of a Nanotubular $ZrO_2$ Layer*

A layer of nanotubular zirconium dioxide ($ZrO_2$-nt) was fabricated by two-stage anodic oxidation of a Zr substrate containing Hf < 4% in a two-electrode cell at a constant voltage of 20 V. A 100 μm-thick Zr foil was preliminarily degreased with acetone, treated with an acid solution in the ratio $HF:HNO_3:H_2O$ = 1:6:20, washed with distilled water and dried in air. The electrolyte was a solution of ethylene glycol containing 5 wt.% $H_2O$ and 1 wt.% $NH_4F$ [42]. During anodization, the anode and electrolyte were maintained at constant temperatures of 10 and 20 °C, respectively. The duration of the primary and secondary anodizing amounted to 5 min. After primary anodization, the obtained oxide layer was etched with the above acid solution.

### *Creating Zr/$ZrO_2$-nt/Au Sandwich Structures*

50 nm-thick Au contacts were deposited onto the as-grown $ZrO_2$-nt layer by magnetron sputtering of gold through a mask mounted on a combined system Q150T ES, Quorum Technologies. Within the sequence-linked operations of the technological cycle specified, more than 100 independent Zr/$ZrO_2$-nt/Au sandwich structures were produced in such a way.

### *Characterization of Zr/$ZrO_2$-nt /Au Sandwich Structures*

The morphology of the as-grown oxide layer was analyzed using a SIGMA VP scanning electron microscope (SEM), Carl Zeiss. The obtained SEM images were processed by a SIAMS 800 automatic system. Imaging the fabricated sandwich structures was carried out using an Axio CSM 700 confocal optical microscope, Carl Zeiss. Built-in standard tools of the Axio CSM 700 Software program provided analyzing the pictures. Figure 1*a* shows an optical microscope image for a series of memristor structures under study. It can be inferred that all the fabricated memristors have a diameter of 140 ± 5 μm and surface roughness parameters such as arithmetical mean deviation of the assessed profile $R_a$ = 250 nm and root mean squared $R_q$ = 350 nm.

Structural characterization of as-grown oxide layer was performed by Rigaku Min-iFlex 600 diffractometer with a copper anode using the Rietveld method. The scanning speed was 0.3 °/min with step of 0.02 °. The diffraction data analysis was carried out in the SmartLab Studio II software.

*Recording Current-Voltage Characteristics*

The measurements of *I-V* characteristics of the fabricated Zr/$ZrO_2$-nt/Au memristors were taken using a PXIe-4143 NI modular controlled power supply unit and a Cascade Microtech MPS 150 microprobe station [50]. The Zr substrate was grounded, and the harmonic voltage *U(t)* with a frequency of 0.01 Hz and amplitude $U_{max}$ = 4 V was applied to the Au contact (see Figure 1*b*).

Before measuring the *I-V* curves, each as-grown memristor was subjected to electroformation (EF). This procedure was implemented by applying a positive or negative polarity voltage *U(t)* that harmonically varied from 0 to $U_{max}$. The current flowing through the structure was limited to ± 0.1 mA (see Figure 1*c*,*d*, EF line). Once the EF procedure finished, the memristors saved *LRS*. The subsequently resistive switching was carried out both in bipolar and unipolar modes.

In bipolar mode, a positive polarity signal of *U*(*t*) first switched the memristors from *LRS* to *HRS* (*LRS* → *HRS*), which corresponded to the "Reset" operation. Then, a negative polarity signal of *U(t)* returned them in the initial state (*HRS* → *LRS*). Such an operation was identified as the "Set" one (see Figure 1*c*).

In unipolar mode, only a positive polarity voltage *U(t)* brought about switching the memristors. First, the "Reset" operation and then the "Set" operation were implemented (see Figure 1*d*). The "Reset" and "Set" operations successively performed comprise one complete cycle of the resistive switching of the memristor. Figure 1*e* sketches a flowchart of the *I-V* recording process. Thus, the *I-V* curves of the fabricated memristors were recorded over 20 full cycles of both resistive switching modes. The electrical resistance of the memristors in *LRS* ($R_{LRS}$) and *HRS* ($R_{HRS}$) was determined from the experimental dependencies of the current passing through the memristor on the applied voltage of $U = \pm$ 0.5 V, *I*(*U*).

So, a schematic representation of the Zr/$ZrO_2$-nt/Au memristor with connected electrical contacts and the order of the *I-V* curve measurements are shown in Figure1*b* – *e*.

*Electric Capacity Measurement*

The electric capacity of the memristors at hand was measured after switching *LRS* → *HRS*. For this, a NI PXI-4072 modular digital multimeter and a Cascade Microtech MPS 150 microprobe station were used.

**Results**

*Structural Characterization*

Figure 2*a* contains SEM images of a synthesized nanotubular $ZrO_2$-nt layer. In the presented image, the oxide layer has a thickness of ≈ 1.7 μm and consists of nanotubes vertically oriented to the substrate. The nanotubes are open on one side (Figure 2*b*), and closed on the side of the Zr-substrate (Figure 2*c*).

Figure 2*d* displays a histogram and a curve of the normal internal-diameter-distribution (solid black line in the Figure) for a series of $10^3$ nanotubes. Analyzing the obtained images showed that the synthesized oxide layer consists of an ordered array of nanotubes whose average values of inner and external diameters are (55 ± 7) and (75 ± 10) nm, respectively. The average wall thickness of nanotubes amounts to (10 ± 5) nm.

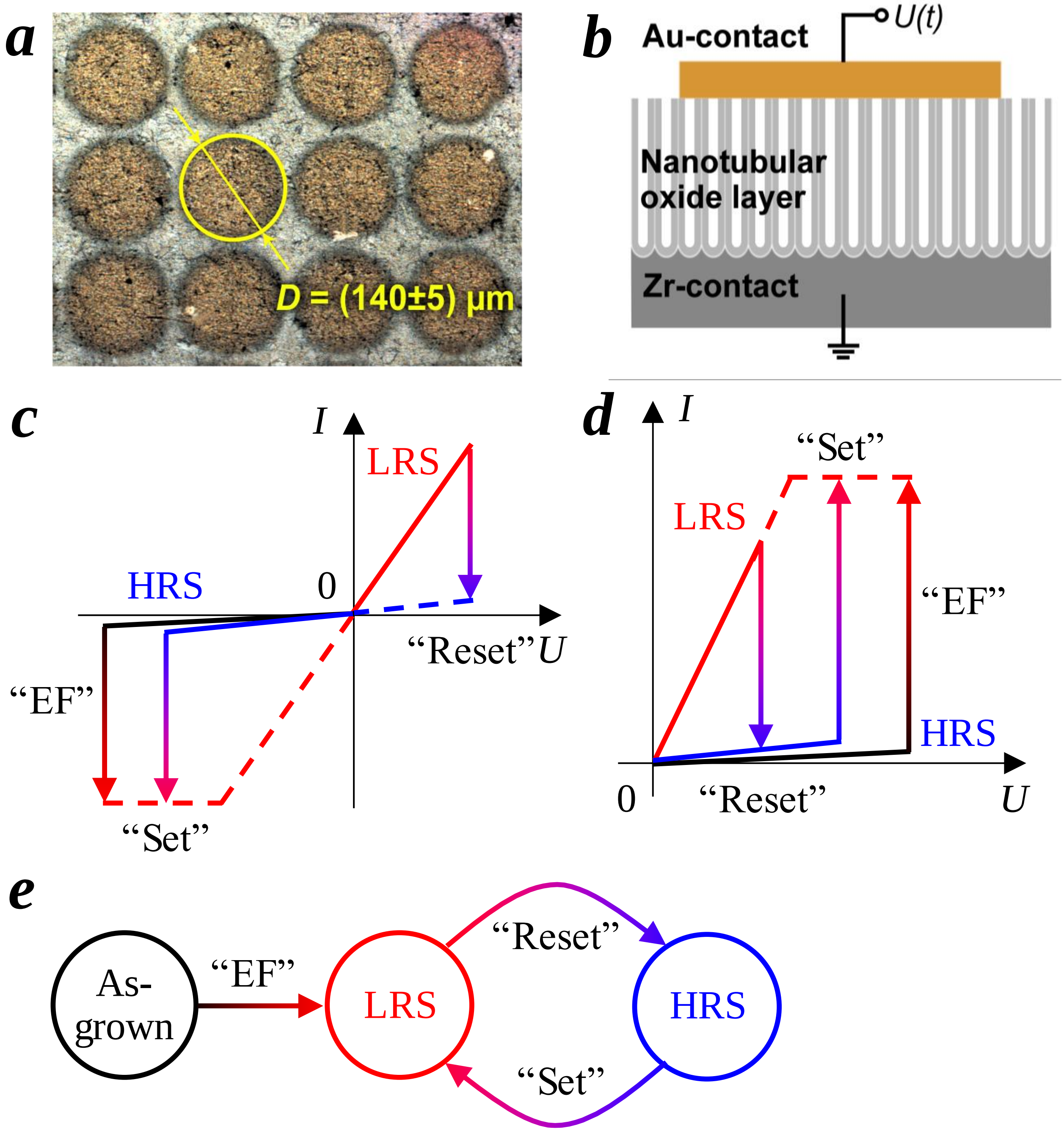

**Figure 1.** Images of fabricated Zr/$ZrO_2$-nt/Au memristors and the *I*-*V* measurement sequence. (*a*) Optical image of Au contacts on the surface of the oxide layer. (*b*) Schematic representation of the structure of a single memristor with connected electrical contacts. (*c*, *d*) Schematic *I*-*V* curves of memristors operating in bipolar and unipolar switching modes, respectively. (*e*) The sequence of changes in the resistive states of memristors.

Figure 3 shows the XRD pattern of $ZrO_2$-nt layer. When the background signal (see Fig. 3, dash line) is taken into account, it can be seen that there are a halo (Fig. 3, gray filling) and two broad peaks with maxima near 2θ = 30 and 50 ° (Fig. 3 , violet filling) on the experimental curve. The halo is due to the amorphous phase, the peaks are related to the crystal structure of the $ZrO_2$-nt sample. The observed peaks can be presented by reflexes superposition from tetragonal (t) (scattering angles 2*θ* = 30.86, 34.32, 36.49, 43.29, 50.96, 52.57, 59.32, 62.21, 64.30, 72.33, 77.54, 78.14, 83.22°, see Fig. 3, green pattern) and monoclinic (m) (at 2*θ* = 18.41, 25.47, 27.54, 35.25, 35.326 37.32, 39.81, 49.94, 50.94, 52.32, 53.80, 55.55, 57.35, 57.67, 61.67, 65.97, 70.29°, see Fig.

3, red pattern) phases. Using standard analysis, it was evaluated that the nanotubes composition is characterized by mixture of t- (43%), m- (32%) and a-ZrO2 (25%) phases.

It is well known that for bulk zirconia the phase transition of the monoclinic structure to the tetragonal one takes place at 1170 °C [51]. However, the data obtained are in satisfactory agreement with studies of phase transformations in the low-temperature sintering process (200 – 350 °C) during the crystallization of an amorphous $ZrO_2$ nanopowder performed by decomposition of zirconium carbonate [52]. The crystallization [52] proceeds with the formation of predominantly t-phase with 36 ± 6 nm grain size and the appearing of m-$ZrO_2$ (15 – 20 %). In our case, the dominance of the tetragonal symmetry may exhibit the presence of structural features < 30 nm – for example, the wall thickness of the synthesized zirconia nanotubes is near 10 nm, see *Structural Characterization section*.

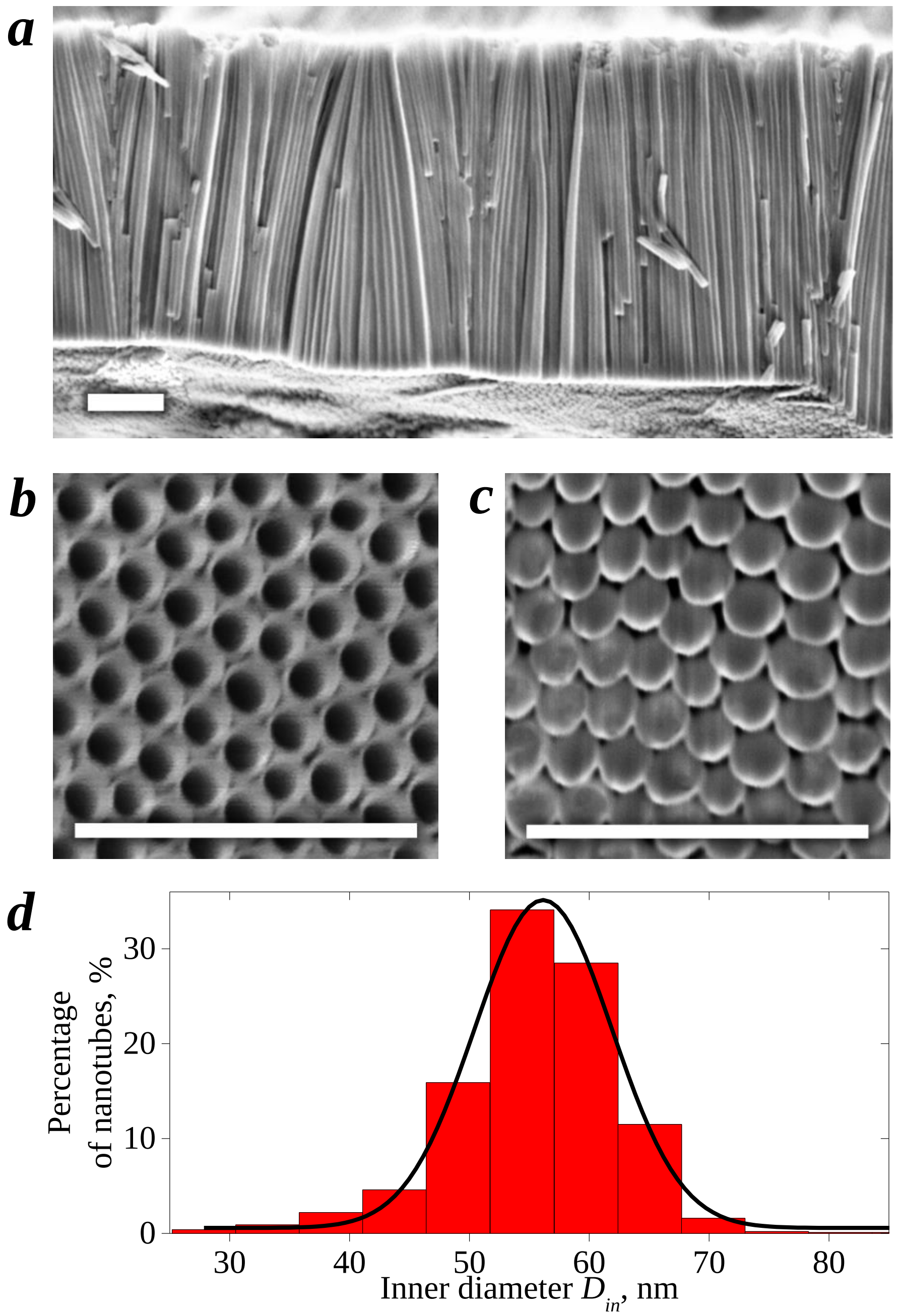

**Figure 2.** SEM characterization of an anodized $ZrO_2$-nt layer. (*a* – *c*) Cross-section, top view (from the side of Au contacts deposited) and bottom view (from the side of the Zr substrate) of the synthesized oxide layer, respectively. (*d*) A histogram of the internal-diameter distribution of nanotubes. The black solid line is a normal distribution.

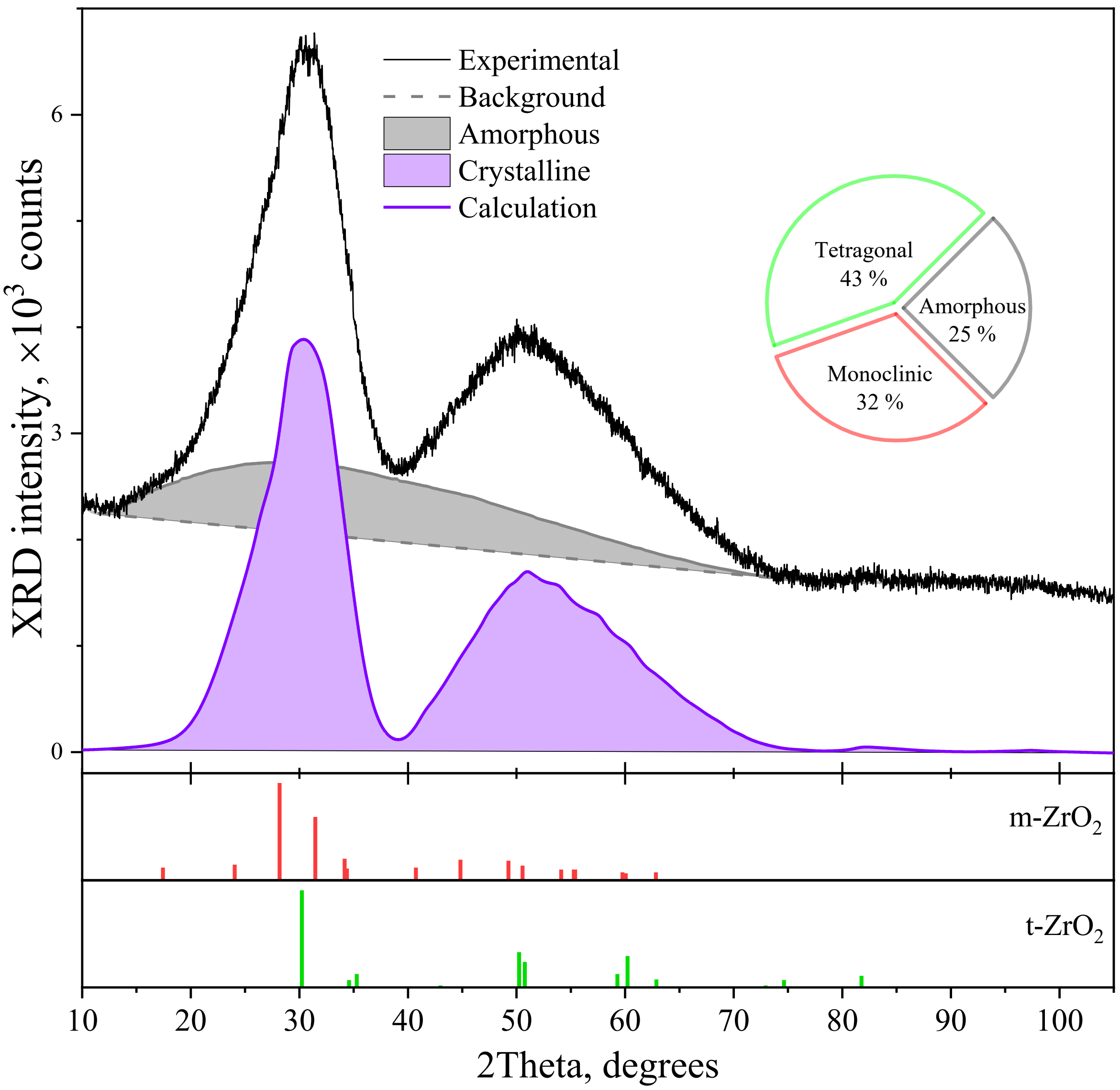

**Figure 3.** XRD patterns and phase composition analysis for synthesized $ZrO_2$-nt layer.

*Current-Voltage Characteristics*

Figures 4*a*, *b* presents the experimental *I-V* curves of the memristors investigated for 20 cycles of bipolar and unipolar switching, respectively. Vertical arrows indicate the instants of a sharp change in current through memristors due to the transition of the structure between resistive states ("Reset" and "Set" operations). Some *I-V* characteristics contain short-term surges and jumps. This can be explained by a change in the electrical resistance of the structure when *U* slightly increases. It is clear from the EF curves that the as-grown Zr/$ZrO_2$-nt/Au structure exhibits characteristics close to those being in *HRS*. A further increase in applied voltage leads to the primary formation of CFs at $U_{EF} \approx$ -2.4 and 4V for bipolar and unipolar switching modes, respectively. Subsequently, the "Reset" operation is realized at voltages of $U_{RES}$ = 0.6 ± 0.1 and 1 ± 0.1 V, and the "Set" operation takes place at voltages of $U_{SET}$ = -2 ± 0.3 and 3.5 ± 0.5 V for bipolar and unipolar modes, respectively.

Thus, the formation of CFs in the oxide layer shunts the Schottky barrier at the Au/$ZrO_2$ interface [25, 35] and eliminates the unidirectional conductivity effect in the as-grown Zr/$ZrO_2$-nt/Au structure [42].

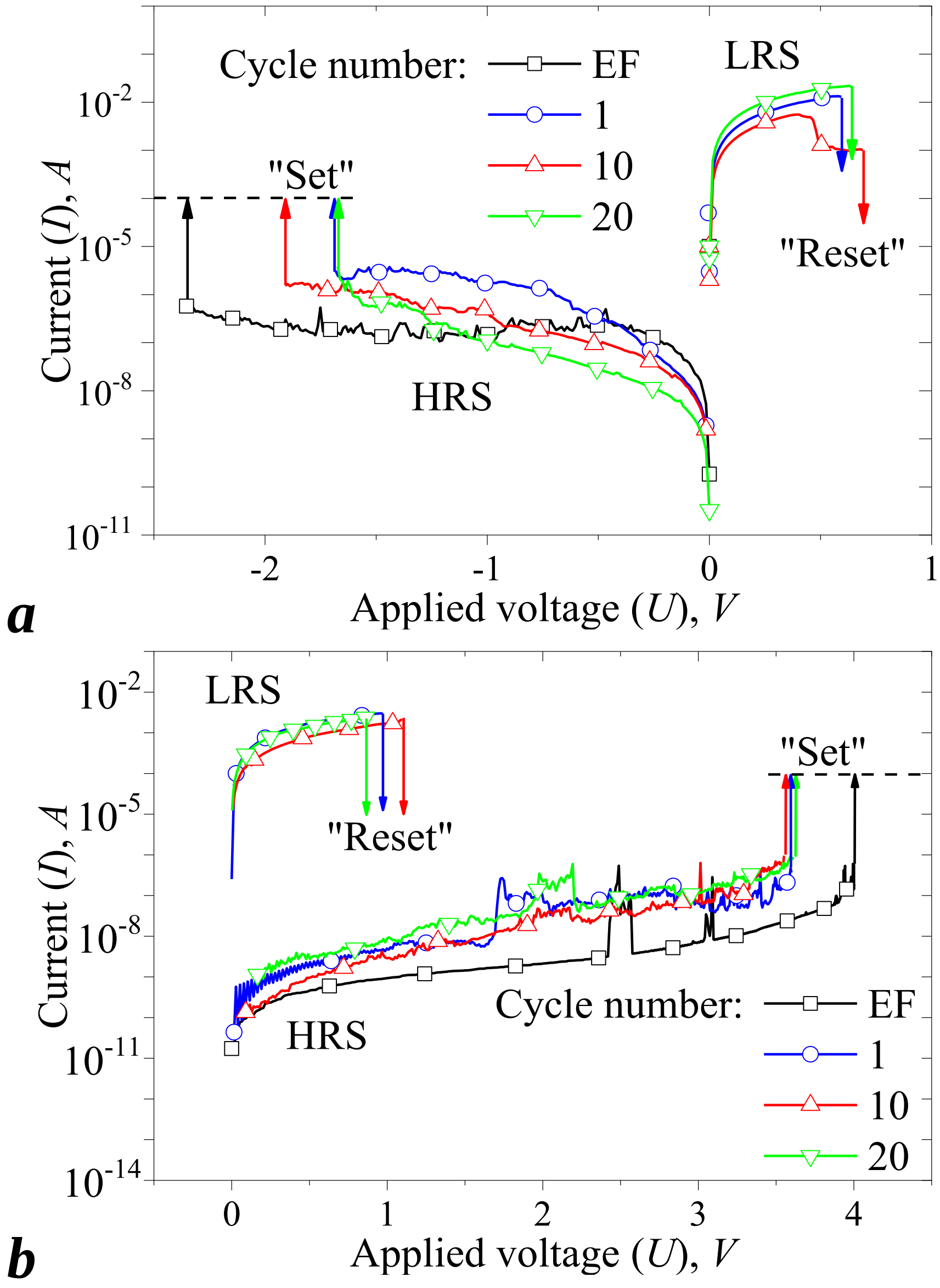

**Figure 4.** Current-voltage characteristics of Zr/$ZrO_2$-nt/Au memristors in (*a*) bipolar and (*b*) unipolar resistive switching modes, respectively. Vertical arrows indicate a change in the resistive state of the memristors. Dotted lines denote the limitation level of current passing through the memristors during their *HRS* → *LRS* switching. EF is electroforming curves. Digits are numbers of complete measurement cycles.

## Discussion

For discussing the findings secured, all the experimental *I-V* characteristics (see Figure 4) were plotted in the following coordinates: surface current density *J* depending on an electric field *E*, with the oxide layer thickness $d = 1.7$ μm and memristor's Au contact area $S = 1.5\times10^4$ $\mu m^2$ taking into account. An analysis of the *J*(*E*) curves accounts for the geometric dimensions of the memristor structures and compares their resistive switching characteristics with those in other papers. Table 1 summarizes the switching mode parameters and morphological features of the Zr/$ZrO_2$-nt/Au structure in comparison with independent data for memristors based on amorphous [22, 23, 24, 26, 36], polycrystalline [16, 17, 20, 29, 31, 33, 53] and nanocomposite [34] $ZrO_2$ layers.

### *Analysis of C-V Curves*

Figures 5*a*, *b* illustrates changes in the values of the electric field strength ($E_{SET}$ and $E_{RES}$) and resistance ($R_{LRS}$ and $R_{HRS}$) for 20 complete cycles of resistive switching of the memristors in bipolar

and unipolar modes, respectively. The experimental values of $E$ for the switching memristors are in the ranges $E_{SET}$ = -10 – -14 and 17 – 23 kV/cm and $E_{RES}$ = 2.9 – 4.1 and 5.4 – 6.4 kV/cm for bipolar and unipolar modes, respectively. It should be emphasized that $E_{SET} \approx (3 - 4) E_{RES}$ for both switching modes of the memristors tested. At the same time, the scatter of the $E_{SET}$ and $E_{RES}$ values for the memristors may take place because of the variation of the $ZrO_2$-nt layer thickness within $1.7 \pm 0.1$ μm and the uncontrolled creation of a multitude of conductive filaments during electroforming.

From Table 1, it is seen that the primary formation of filaments and further resistive switching in the synthesized Zr/$ZrO_2$-nt/Au is realized at lower $E$ values despite a thicker oxide layer compared to the memristor structures presented. This appears to be caused by several factors, for example, firstly, due to a high concentration of oxygen vacancies in the anodized $ZrO_2$. So in [54], the X-ray photoelectron spectroscopy method demonstrates a high photocatalytic activity of oxygen vacancies in nanotubular $ZrO_{2-X}$ obtained by the anodization method.

Secondly, the oxide layer of the memristors has the nanotubular morphology. Simulating the distribution of the electric field strength throughout the nanotubular structure confirms the emergence of increased strength regions in the base of the nanotubes, which determines their growth during anodization (see the example of $TiO_2$ in [55]). Besides, the influence of surface roughness of the Au/$TiO_2$ interface for memristor structures based on a 20 nm thick $TiO_2$ film synthesized by magnetron sputtering was explored in [56]. The authors showed that the values of memristor resistive switching voltages diminish with increasing the roughness in the range of $Rq$ = 4.2 – 13.1 nm. This is due to the fact that the sharp roughness peaks heighten the electric field strength. In this case, the formation of CFs occurs in the active layer at lower voltages [56]. Note that in our case, the value of the $Rq$ roughness parameter is noticeably larger (see Section Results. Structural characterization). Thus, the electric and physical characteristics of memristor structures can be improved by the synthesis of thinner nanotubular oxide layers.

Thirdly, the content of Hf impurities in the active layer also can be among the aforementioned factors. This is confirmed by the results of the performed quantitative chemical analysis. Independent data calculated by the density-functional method testifies that the incorporation of isovalent Hf into $ZrO_2$ contributes to a slight decrease in the $V_O$-formation energy in the oxide and the improvement of memristor characteristics [27].

It is evident from Figure 5 and Table 1 that the structure resistance in *LRS* is hundreds of Ω, and tens and hundreds of MΩ in *HRS*. It is worth noting that the values of $R_{HRS}$ and $R_{LRS}$ coincide with similar parameters for $ZrO_2$-based memristor structures despite the significantly larger thickness of the $ZrO_2$-nt layer. There are a lot of papers that contain the fact that the $R_{HRS}/R_{LRS}$ resistance ratios in different states lie in the range of 10 – $10^5$ (see Table 1). In our case, $r = R_{HRS}/R_{LRS} > 37 \cdot 10^4$ and $8 \cdot 10^4$ for unipolar and bipolar switching modes, respectively. The experimental values are comparable with those of the best memristor structures based on thin amorphous $ZrO_2$ films obtained by electron beam evaporation [22] and subsequent implantation of Zr [36] or Cu [24] ions, see Table 1.

Having compared the obtained parameters of the resistive switching of thin $ZrO_2$-layer-based Zr/$ZrO_2$-nt/Au structure memristors, the following can be stated: in bipolar mode, lower values of the electric field strength contribute to switching between *HRS* and *LRS*. In the process, lower values of $r$ are recorded. The values of the experimentally measured switching parameters of the structures studied allow one to claim that anodized $ZrO_2$-based nanotubular layers are promising to use as memristor memory cells.

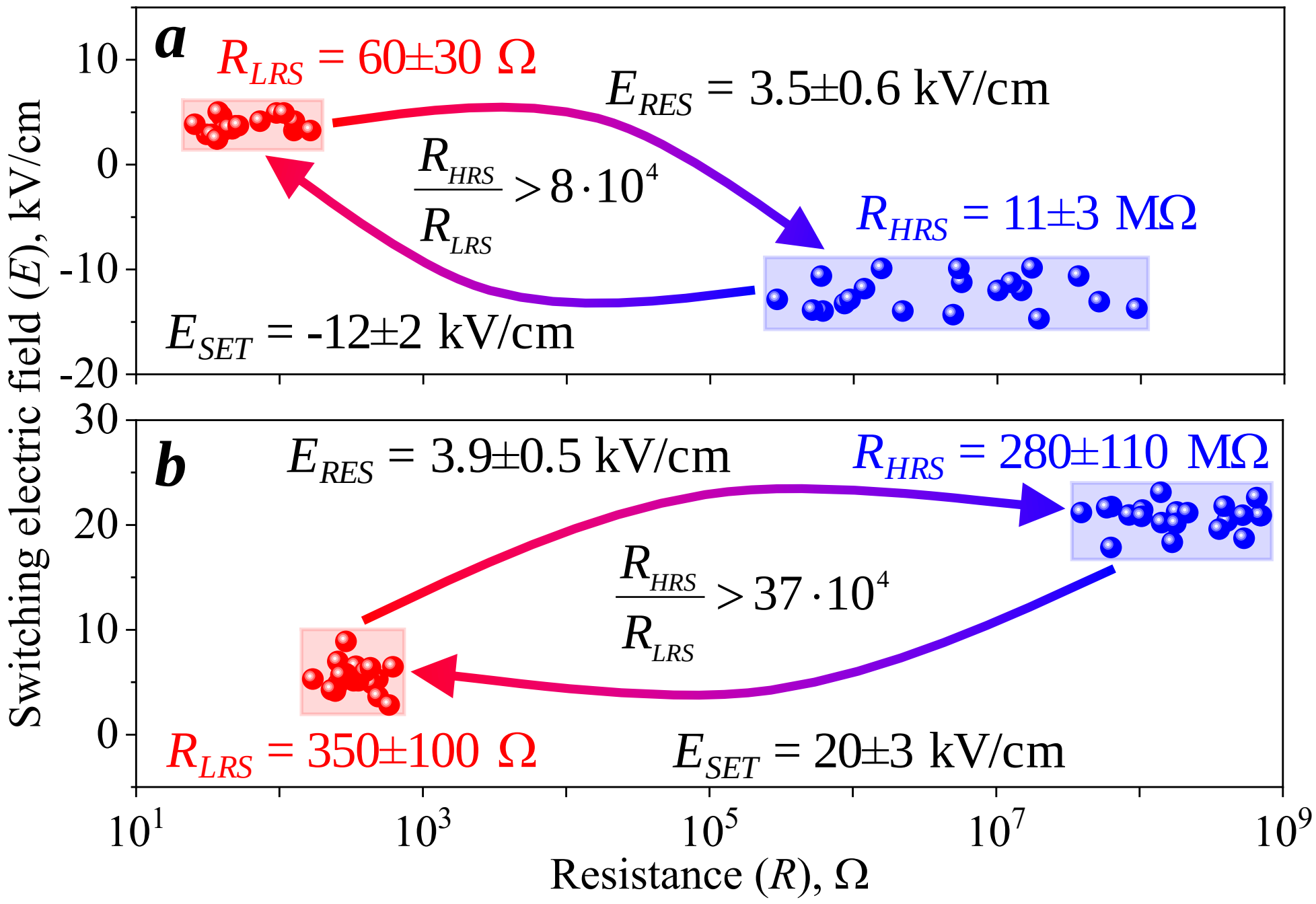

**Figure 5.** Change in electric field strength ($E$) and resistance ($R$) for 20 complete cycles of resistive switching of Zr/$ZrO_2$-nt/Au memristors for (*a*) bipolar and (*b*) unipolar modes. Rectangles indicate the regions of variation of the evaluated characteristics.

**Table 1.** Switching electric field (*E*) and electrical resistance (*R*) values for memristors structures based on the $ZrO_2$ layers with different morphologies and thickness (*d*)

| № | Mode | Operation, state | *E*, kV/cm | *R*, Ω | $r = R_{HRS}/R_{LRS}$ | Conduction mechanism and/or behavior | Memristor: structure, size, CFs type | Active layer: thickness, morphology, synthesis technique | Reference |
|---|---|---|---|---|---|---|---|---|---|
| 1 | – | EF, as-grown | 23.5<br>-14 | $10^9$<br>$1.7 \cdot 10^7$ | – | Ohm[I] | $Zr/ZrO_2$-nt/Au, Ø 140 μm ($1.4 \times 10^4$ $\mu m^2$), $V_O$ | 1.7 ± 0.1 μm, nanotubular PC[VIII] layer, anodization | This work |
| | Unipolar | Set, *HRS* | 20±3 | $(28 \pm 11) \cdot 10^7$ | > $37 \cdot 10^4$ | SCLC[II] | | | |
| | | Reset, *LRS* | 5.9±0.6 | 350±100 | | Ohm | | | |
| | Bipolar | Set, *HRS* | -12±2 | $(11 \pm 3) \cdot 10^6$ | > $8 \cdot 10^4$ | SCLC | | | |
| | | Reset, *LRS* | 3.5±0.6 | 60±30 | | Ohm | | | |
| 2 | – | EF | 625 | No data | – | No data | $Zr/ZrO_2/Au$, Ø 500 μm or 25×25 $\mu m^2$ | 40 nm, no data, anodization | [21, 49] |
| | Unipolar | Set, *HRS* | 225…725 | $(5…260) \cdot 10^4$ | > $10^2$ | | | | |
| | | Reset, *LRS* | 75…325 | 26…130 | | | | | |
| 3 | – | EF | 415–580 | No data | – | No data | Pt/ZrO/$p^+$-Si, $9 \times 10^{-6}$ $cm^2$, $Zr^+$ | 120 nm, PC (*t*[IX] and *m*[X]) film, magnetron sputtering | [16] |
| | Unipolar | Set, *HRS* | 135 | $1…10^4$ | > 3 | SE[III], Ohm | | | |
| | | Reset, *LRS* | 90 | | | Ohm | | | |
| 4 | Bipolar | Set, *HRS* | ±430 | $10^8$ | > $3 \cdot 10^4$ | SCLC | $Au/Cr/ZrO_2:Zr^+/n^+$-Si, from 50×50 $\mu m^2$ to 800×800 $\mu m^2$ | 70 nm, Zr-implanted amorphous film, EBE[IV] | [36] |
| | | Reset, *LRS* | | $3 \cdot 10^3$ | | | | | |
| 5 | – | EF, as-grown | 1400 | $10^9$ | No data | No data | $Au/Cr/ZrO_2:Au/n^+$-Si, from 100×100 $\mu m^2$ to 1000×1000 $\mu m^2$ | 70 nm, Au-implanted amorphous film, EBE | [26] |
| | Unipolar | Set, *HRS* | 1100 | $(2–500) \cdot 10^6$ | > 66 | SCLC | | | |
| | | Reset, *LRS* | 290 | $(1–300) \cdot 10^2$ | | VRH[V] | | | |
| | Bipolar | Set, *HRS* | ±570 | $(2–500) \cdot 10^6$ | | SCLC | | | |
| | | Reset, *LRS* | | $(1–300) \cdot 10^2$ | | VRH | | | |
| 6 | – | EF | No data | > $10^9$ | – | No data | $Ag/ZrO_2:Cu$/ Pt, 0.5×0.5 $\mu m^2$ | 40 nm, amorphous layer with Cu nanocrystal, EBE | [24] |
| | Bipolar | Set, *HRS* | 125 | $(1–300) \cdot 10^7$ | > $3 \cdot 10^4$ | | | | |
| | | Reset, *LRS* | -50 | 200…320 | | | | | |
| 7 | Bipolar | Set, *HRS* | 420 | $9\cdot10^4$ | $> 2\cdot10^3$ | SE, Ohm | Al/$ZrO_2$/Al, Ø 80 μm | 60 nm, amorphous stoichiometric film, magnetron sputtering | [22] |
| | | Reset, *LRS* | -80 | 40 | | Ohm | | | |
| 8 | Bipolar | Set, *HRS* | > 600 | $10^5$–$10^6$ | $> 10^2 \ldots 10^3$ | P-F[VI] | ITO/$ZrO_2$/Ag, 60×60 $\mu m^2$, Ag | 20…50 nm, solid PC (c[XI]) film, sol-gel process | [17] |
| | | Reset, *LRS* | < -1400 | $10^3$ | | Ohm | | | |
| 9 | – | EF | 850 | No data | – | No data | Cu/$ZrO_2$/$TiO_2$/Ti, no data | 50 nm, solid amorphous film, EBE | [23] |
| | Bipolar | Set, *HRS* | 1000 | $10^6$ | $> 10^2$ | | | | |
| | | Reset, *LRS* | < -1000 | $10^4$ | | | | | |
| 10 | – | EF | 1375 | No data | No data | No data | Pt/Ti/$ZrO_2$/Pt/Ti 100×100 $\mu m^2$ | 20 nm, solid film, no data | [25] |
| | Bipolar | Set, *HRS* | 250±50 | No data | | SCLC | | | |
| | | Reset, *LRS* | -250±25 | 130 | | Ohm | | | |
| 11 | Unipolar | Set, *HRS* | ≈ 50 | No data | No data | No data | Ag/$ZrO_2$/Pt, Ø 200 μm, Ag | 50 nm, no data, EBE | [18] |
| | | Reset, *LRS* | ≈ 10 | | | | | | |
| 12 | – | EF, as-grown | ≈ 200 | $\approx 10^7$ | – | No data | Cu/$ZrO_2$/Pt, Ø 200 μm, Cu | 50 nm, PC film, EBE | [31] |
| | Unipolar | Set, *HRS* | ≈ 300 | $\approx 10^7$ | > 100 | | | | |
| | | Reset, *LRS* | ~140 | $\approx 1.3\cdot10^4$ | | | | | |
| | Bipolar | Set, *HRS* | -(120±20) | $\approx 10^6$ | | | | | |
| | | Reset, *LRS* | 200±100 | $\approx 10^4$ | | | | | |
| 13 | – | EF | 6000 | No data | – | No data | Ni/$ZrO_2$/TaN, Ø 150 μm | 10 nm, PC film, magnetron sputtering | [53] |
| | Unipolar | Set, *HRS* | $(1.6–2.6)\cdot10^3$ | $3\cdot10^5$ | > 10 | | | | |
| | | Reset, *LRS* | $(6–13)\cdot10^2$ | $2\cdot10^2$ | | | | | |
| 14 | Unipolar | Set, *HRS* | 1050–1800 | $\approx 10^8$ | $> 10^6$ | No data | Cu/$ZrO_2$:Cu/Pt, from 3×3 $\mu m^2$ to 20×20 $\mu m^2$, Cu | 20 nm, Cu-doped layer, EBE | [19] |
| | | Reset, *LRS* | 400–750 | $\approx 10^2$ | | | | | |
| | Bipolar | Set, *HRS* | 1050–1800 | $\approx 10^8$ | | | | | |
| | | Reset, *LRS* | 400–750 | $\approx 10^2$ | | | | | |
| 15 | Bipolar | Set, *HRS* | ≈ 200 | $\approx 10^7$ | $> 10^4$ | Complex | Cu/$ZrO_2$:Ti/Pt, Ti/$ZrO_2$:Ti/Pt, from 50×50 $\mu m^2$ to 800×800 $\mu m^2$, Ti | 70 nm, Ti-implanted PC film, EBE | [20] |
| | | Reset, *LRS* | ≈ 90 | $\approx 10^2$ | | Ohm | | | |
| 16 | – | EF | ≈ 1800 | No data | – | No data | $TaN/ZrO_2/Pt$, $TaN/ZrO_2:Gd/Pt$, $TaN/ZrO_2:Dy/Pt$, $TaN/ZrO_2:Ce/Pt$, Ø 100 μm | (13…25) nm, pure, Gd-, Dy- or Ce-doped PC (t and c) film, sol-gel process | [29, 33] |
| | Bipolar | Set, *HRS* | No data | | > 10 | | | | |
| | | Reset, *LRS* | | | | | | | |
| 17 | Bipolar | Set, *HRS* | ±75 | $11.5 \cdot 10^3$ | > 44 | SCLC | $ITO/ZrO_2:PVP^{VII}/Ag$ $100 \times 100\ \mu m^2$ | 200 nm, nanocomposite layer, spin coating | [34] |
| | | Reset, *LRS* | | 263 | | Ohm | | | |
| 18 | Bipolar | Set, *HRS* | ≈ (1.5…1.8) $\cdot 10^4$ | > 130 | > 32 for $Ag/ZrO_2/ITO$, > 1.7 for $Ag/ZrO_2/Ag$ | SCLC | $Ag/ZrO_2/Ag$, $Ag/ZrO_2/ITO$, no data, $Ag^+$ | 100 nm, PC (m) film, magnetron sputtering | [57] |
| | | Reset, *LRS* | ≈ (0.7…1.5) $\cdot 10^4$ | < 115 | | Ohm | | | |

[I] Ohm is Ohmic conduction
[II] SCLC is space-charge-limited conduction
[II] SE is Schottky emission
[IV] EBE is electron beam evaporation
[V] VRH is Mott variable range hopping
[VI] P-F is Poole-Frenkel emission
[VII] PVP is Poly (4-vinylphenol)
[VIII] PC is polycrystalline
[IX] *t* is the tetragonal phase
[X] *m* is the monoclinic phase
[XI] *c* is the cubic phase

*Conduction Mechanisms*

*HRS*

Figure 6 shows the *J(E)* data in double logarithmic coordinates for both resistive switching modes. It is seen that the measured dependencies for *HRS* contain three linear segments with different tangents of inclination angles. In browsing the literature [58, 59, 60], it can be concluded that the above behavior corresponds to the *I-V* characteristics of a dielectric with monoenergetic traps capable of filling with injected electrons. The theory of injection currents reads that such systems can implement the space-charge-limited conduction (SCLC) mechanism [58]. The SCLC model succeeds at describing carrier transport through a dielectric layer by the following equations [58, 59, 60]:

$$J_{Ohm} = \frac{I_{Ohm}}{S} = e\mu n_0 \frac{U}{d} = e\mu n_0 E \, , \quad (1)$$

$$J_{TFL} = \frac{I_{TFL}}{S} = \frac{9}{8}\mu_{eff}\varepsilon\varepsilon_0 \frac{U^2}{d^3} = \frac{9}{8}\mu_{eff}\varepsilon\varepsilon_0 \frac{E^2}{d} \, , \quad (2)$$

$$\mu_{eff} = \mu\theta \, , \quad (3)$$

$$\tau_c = \frac{d^2}{\mu_{eff} U_{tr}} = \frac{d}{\mu_{eff} E_{tr}} \, , \quad (4)$$

$$\tau_d = \frac{\varepsilon\varepsilon_0}{en\mu_{eff}} \, , \quad (5)$$

$$E_{TFL} = \frac{U_{TFL}}{d} = \frac{eN_t d}{2\varepsilon\varepsilon_0} \, , \quad (6)$$

where $J_{Ohm}$ is Ohm's law current density, A/cm$^2$; $I_{Ohm}$ is Ohm's law current, A/cm$^2$; $S$ is the memristor area, cm$^2$; $e$ is the electron charge, C; $\mu$ is the electronic drift mobility, cm$^2$/(V·s); $n_0$ is carrier concentration in thermal equilibrium, cm$^{-3}$; $U$ is applied voltage, V; $d$ is the thickness of oxide layer, cm; $E$ is applied electric field, V/cm; $J_{TFL}$ is trap-filled limit current density, A/cm$^2$; $I_{TFL}$ is trap-filled limit current, A; $\mu_{eff}$ is effective electron mobility, cm$^2$/(V·s); $\varepsilon$ is the static dielectric constant; $\varepsilon_0$ is the permittivity in vacuum, F/cm; $\theta$ is the ratio of the free carrier density to total carrier density; $\tau_c$ is carrier transit time, s; $U_{tr}$ is transition voltage, V; $E_{tr}$ is transition electric field, V/cm; $\tau_d$ is dielectric relaxation time, s; $n$ is the concentration of the free carriers in the oxide, cm$^{-3}$; $E_{TFL}$ is the trap-filled limit electric field, V/cm; $U_{TFL}$ is the trap-filled limit voltage, V; $N_t$ is the trap density, cm$^{-3}$.

In our case, electrons are the charge carriers in the $ZrO_2$-nt layer, and oxygen vacancies $V_O$s are traps. From now on, when discussing experimental data within the SCLC model [58, 59, 60], we will adhere to this statement.

For $E < E_{tr}$, the $J(E)$ dependence is linear ($J_{Ohm} \sim E$) and obeys Ohm's law in Eq. (1). Such a regime for $ZrO_2$ is typical of an electrically quasi-neutral state that corresponds to the early stage of filling the $V_O$s at weak injection $N_{t0} \to 0$.

In the case of $E = E_{tr}$, the number of injected and free electrons in the active memristor layer coincides ($n = n_0$), and the transit times (see Eq. (4)) and electron relaxation (see Eq. (5)) are equal, i.e. $\tau_c = \tau_d$. Moreover, $\Theta \to 1$ since the number of filled traps is $N_{t,0} \to 0$.

For $E_{tr} < E < E_{TFL}$, the condition of strong injection holds. In the concerned range of $E$, the traps continue filling up, and the dependence becomes quadratic ($J_{TFL} \sim E^2$). As $E$ grows, the concentration of free injected electrons increases, which in turn leads to a shift in the position of the Fermi quasi-level to the position of the level of traps in the bandgap.

For $E \geq E_{TFL}$, the current flowing through the structure abruptly goes up, and the double logarithmic dependence $J(E)$ has a slope $k \geq 3$ ($J \sim E^k$). This fact evidences the complete filling of traps and the possibility of freely moving the injected electrons in the $ZrO_2$ layer. Thus, the $E_{TFL}$ strength matches the $V_O$ levels completely filled (see Eq. (6)) and (or) the coincidence of the positions of the Fermi quasi-levels and monoenergetic $V_O$ levels in the bandgap [58, 59]. Then, a very strong injection is the reason for CFs to form in the electric field range in question.

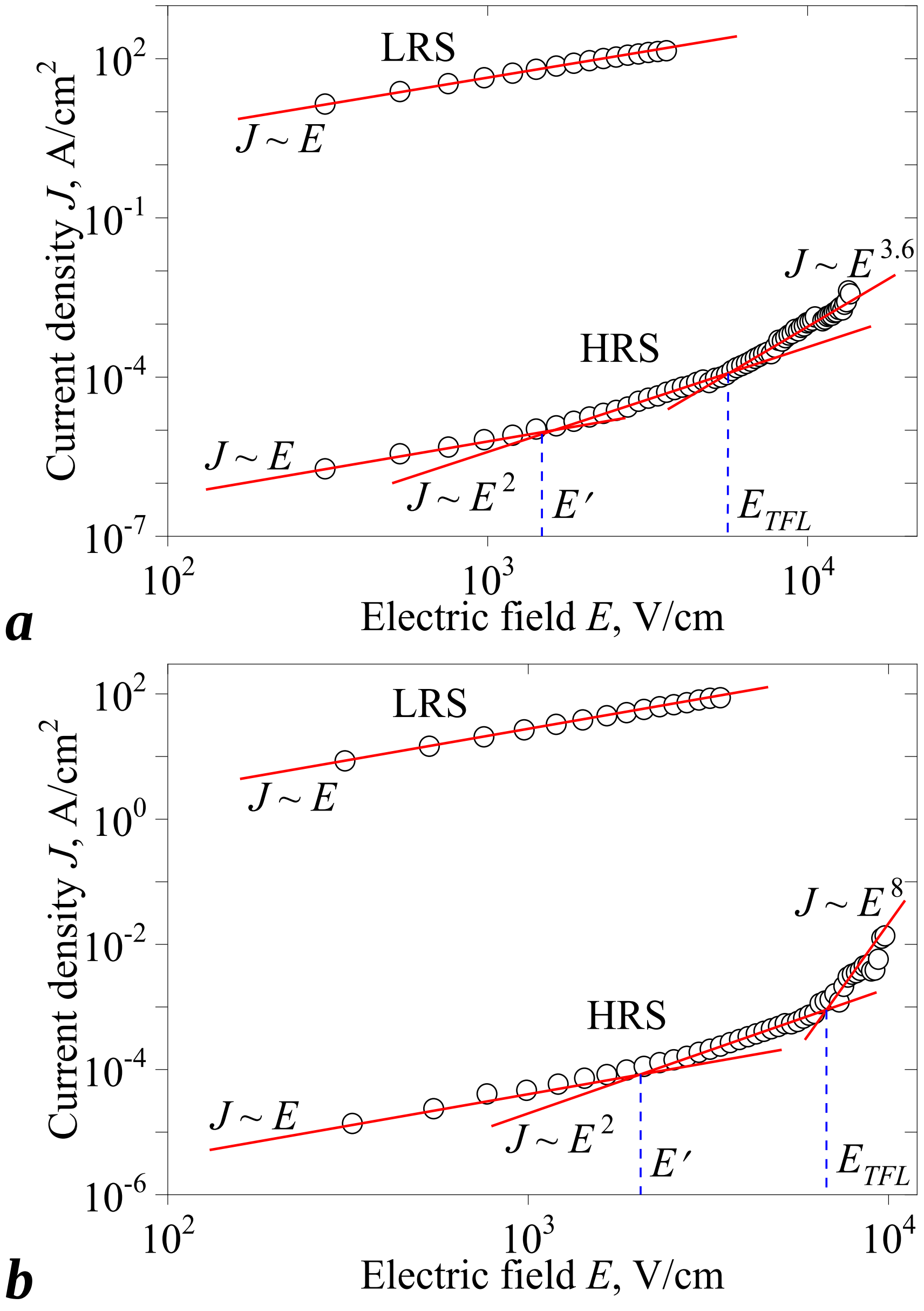

**Figure 6.** Dependencies of current density ($J$) on strength ($E$) in double logarithmic coordinates for bipolar ($a$) and unipolar ($b$) switching modes. Symbols are the experimental data. Solid lines represent a linear approximation.

To evaluate the parameters of the $ZrO_2$-nt layer using Eq. (1) – (6), we measured the electric capacitance of memristors in *HRS*, which amounted to $C$ = 1 pF. In the case of the geometry of a flat capacitor, we obtain the relative permittivity of $\varepsilon = 20 \pm 2$. This value is quite consistent with $\varepsilon$ = 14 – 40 for continuous thin zirconium films [28, 61, 62] and coincides with the estimate of $\varepsilon$ = 21 for an anodized $ZrO_2$ layer synthesized in a galvanostatic mode at room temperature in an electrolyte of ammonium tetraborate [63]. Here, the morphology of the oxide layer is not covered.

The electric and physical characteristics of the $ZrO_2$-nt layer in *HRS*: $\mu_{eff} \approx \mu = (0.1 - 12)\cdot 10^{-3}$ $cm^2\cdot(V\cdot s)^{-1}$ $N_t = (1 - 3)\cdot 10^{15}$ $cm^{-3}$, $\tau_c$ = 0.3 – 1.3 ms and $n_0 = (1 - 3)\cdot 10^{14}$ $cm^{-3}$ were calculated according to Eq. (1) – (6). It is seen that the mobility values obtained at room temperature are comparable with independent estimates of $\mu = 23\cdot10^{-3}$ [64] and $5.8\cdot10^{-3}$ $cm^2\cdot(V\cdot s)^{-1}$ [65] at 900 °C for single-crystal $ZrO_2$ stabilized $Y_2O_3$.

It is important to note that, in a number of independent works, both theory and experiment confirm a high concentration of $V_O$ in various anodized oxides [38,63,66,67,68,69]. For example, when treated at a temperature of 500 °C for 4 h in a reducing atmosphere of 5% $H_2/N_2$, $TiO_2$ nanotubular array increases the concentration of active surface defects, with the charge carrier density becoming equal to $9.86\cdot10^{20}$ $cm^{-3}$ [68]. Large values of $2.5\cdot10^{21}$ $cm^{-3}$ for the $V_O$ concentration in $ZrO_2$ single crystals stabilized by $Y_2O_3$ were previously determined by the electron paramagnetic resonance (EPR) method [65].

The significantly lower value of $N_t = (1 - 3)\cdot10^{15}$ $cm^{-3}$, obtained in the present work can be explained by the presence of $F^-$ ions. It is known that fluorine substitutes for oxygen vacancies in ionic crystals, which leads to a decrease in the concentration of $V_0$ and eliminates allowed capture levels in the bandgap of oxides [70]. X-ray fluorescence and photoelectron spectroscopy methods confirm the presence of fluorine in the samples under study. The fluorine and hafnium contents are found to amount to 2.5 and 1 at. %, respectively.

*LRS*

It is clear from Figure 6 that, for a low resistive state, all the *J*(*E*) dependencies in double logarithmic coordinates are linear with the slope tangent ≈ 1 ($J_{Ohm} \sim E$), which corresponds to Ohm's law. Such a behavior is also typical of memristor structures based on continuous $ZrO_2$ layers in *LRS* (see Table 1) [16, 17, 20, 22, 25, 34]. In addition, the *I-V* curves have a region of switching at $U = 0.4 – 1$ V during the *LRS* → *HRS* transition. In this case, the *G* conductance of the memristor structure stepwise decreases. For example, with increasing *U*, the magnitude of *G* diminishes by a factor to one-half and half-integer values of the quantum of electrical conductance $G_0 = 2e^2/h \approx 77.5$ μS (Figure 7, curve X) in the row: $53G_0 \to 51G_0 \to 48G_0 \to 45G_0 \to 13G_0 \to 9G_0 \to 5G_0 \to 4G_0 \to 3G_0 \to 2G_0 \to 1.5G_0 \to 1G_0 \to 0.5G_0 \to G_{HRS}$.

The resistive switching of various memristor structures as a result of the formation of quantum conductive filaments (QCFs) at room temperature is an experimentally and theoretically confirmed fact [71 and ref. in it]. For example, it is established in [31] that quantum conductive filaments (QCFs) in Cu/ $ZrO_2$/Pt memristors are formed in a controlled manner.

Based on the experimental $R_{LRS}$ values (see Table 1), it can be argued that when performing the “EF” and “Set” operations, tens and hundreds of QCFs are formed and restored between the electrodes in memristors. This corresponds to the formation of one QCF per $10^4$-$10^5$ $ZrO_2$ nanotubes, whereas their number in the memristor is ≈ $3.5\cdot10^6$ pieces. It is safe to say that QCFs fully determine electron transport through the memristor in the *LRS*, and the discreteness of the change in the conductance values leads to the quantization of the mobility values.

Let one QCF have a volume $V_{QCF}$, in which two electrons reside. Then, according to Eq. (1), the mobility of elementary charges in *LRS* is discrete:

$$\mu_{LRS} = \mu_0 \frac{xV_{QCF}}{Sd}, \qquad (7)$$

Where $\mu_0 = ed^2/h \approx 698.7$ $m^2\cdot(V\cdot s)^{-1}$. The data of speculations are consistent with theoretical calculations of electron mobility in multi-walled carbon nanotubes within an ideal multi-box model, for which $\mu$ is quantized and proportional to the square of the nanotube length [72].

Eq.(7) implies that $\mu_{LRS}$ depends on the ratio of the geometric dimensions of the memristor and QCF, as well as the number of filaments formed. So, comparing the volumes of memristor and QCFs ($Sd \to x\,V_{QCF}$), we get that $\mu_{LRS} \to \mu_0$. Thus, a possible increase in electron mobility in $ZrO_2$-nt during the transition of the memristor from *HRS* to *LRS* is a result of the quantization of energy states in one-dimensional channels arising from the motion of charge carriers over QCFs in a dielectric layer. It was previously reported that $\mu$ = 12-13 $cm^2\cdot(V\cdot s)^{-1}$ at temperatures of 425-475K for thin amorphous oxide films in the Al/$ZrO_2$/p-Si structure using the modified Schottky emission model [73].

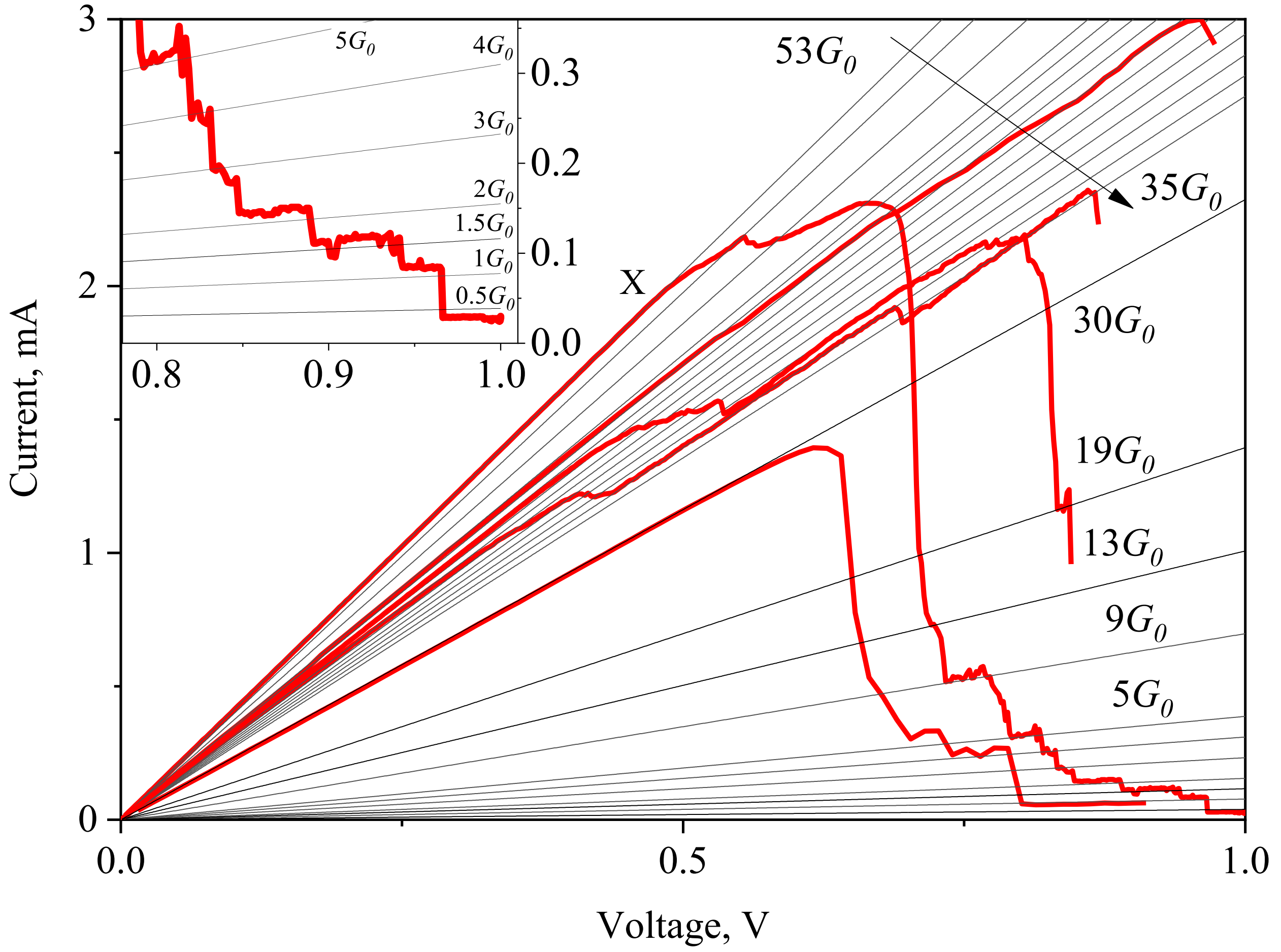

**Figure 7.** Experimental *I-V* characteristics of memristors in *LRS* (solid red lines) and a series of theoretical *I-V* characteristics for CFs with quantum conductance $xG_0$ (thin black lines), where $x$ is the number of QCFs and $G_0 = 2e^2/h \approx 77.5$ μS is the quantum of electrical conductance.

It should be underscored that the generated QCFs in the $ZrO_2$-nt layer are stable even at room temperature and retain their conducting state when $E$ is turned off. Subsequently, during the "Reset" operation, QCFs are destroyed. Next, we explain the mechanisms of electroforming and resistive switching of memristors with the participation of $V_O^{2+}$ oxygen vacancies, $O^{2-}$oxygen and $F^-$ fluorine ions in these processes.

*Resistive Switching Mechanisms*

Figure 8 schematically sketches the processes that occur in nanotubes in the bipolar memristive switching mode. The as-grown nanotube whose fragment is outlined by a dark frame has increased non-stoichiometry in oxygen at the $ZrO_2$-nt/Zr interface [38, 66, 74]. Independent data on the $ZrO_2$-nt layer, obtained by transmission electron microscopy and energy dispersive X-ray analysis, show that the content of $F^-$ ions is also high in this region [74].

When electroforming (see Figure 8, operations EF), several complementary processes can proceed in the Zr/$ZrO_2$-nt/Au structure at hand. Firstly, positively charged oxygen vacancies $V_O^{2+}$ migrate outwards and towards the Au-contact in an applied electric field at $E > 0$ and $E < 0$, respectively. The aforementioned vacancy centers have a positive charge relative to the crystal lattice, and their motion appears to obey the exchange mechanism in the anion sublattice due to the diffusion of $O^{2-}$[75] and $F^-$ ions along the grain boundaries of the nanotubes in an external electric field.

Secondly, the motion of negatively charged $F^-$ and/or $O^{2-}$ ions outwards and towards the Au contact is initiated at $E > 0$ and $E < 0$, respectively. In this case, extra oxygen vacancies emerge in the $ZrO_2$ nanotubes. It is known that the mobility of fluorine in anodized oxides is almost twice that of oxygen [76, 77]. Consequently, the transition of F from the O position to the $F_i^-$-interstice in an external electric field results in $V_O$-forming as a most probable process. The latter proceeds according to the $F_O^+ \rightarrow V_O^{2+} + F_i^-$ mechanism [70]. It should be underscored that the diffusion of O and the formation of voids at the "active layer/electrode" interface during the EF process were experimentally confirmed by studying the distribution of chemical elements in a $Pt/TiO_2/Pt$ memristor using two-dimensional energy-dispersive X-ray spectroscopy [78].

Thus, the QCFs are formed upon locally reaching a certain threshold concentration of $V_O$ in the $ZrO_2$-nt layer, and the memristor passes into the *LRS* state (see Figure 8, a red frame). It is worth emphasizing that the EF process for memristors at $E < 0$ is less energy-consuming as compared to that at $E > 0$ since it proceeds at lower $E_{EF}$ values (see Table 1).

Further, resistive switching is possible to implement after the mandatory EF procedure. So when placing the memristor being in the *LRS* into an electric field *E*, current flows through the QCFs formed. Once heated up due to the Joule heat, they are locally destroyed as a result of diffusion of $V_O$, F and/or O (see Figure 8, Reset operation). Chances are, the QCFs collapse, first of all, inside the nanotube walls far from the metal contacts, and the memristor passes into the *HRS* state (see Fig. 8, a blue frame). Subsequently, the applied field *E* initiates the diffusion process of $V_O$, F, and/or O in locally destroyed areas of the QCFs (see Figure 8, operation Set). When restored, the $V_O$ chains make the memristor go into the *LRS* state. As it becomes clear from Table 1, $E_{SET} > E_{RES}$ for both switching modes of the memristors tested. Thus, partial destruction of the QCFs in the active layer requires less energy than their restoration. The similar behavior of the $Zr/ZrO_2$-nt/Au structure is consistent with a number of independent works [16, 21, 22, 24, 26, 49] for continuous-$ZrO_2$ layer-based memristors.

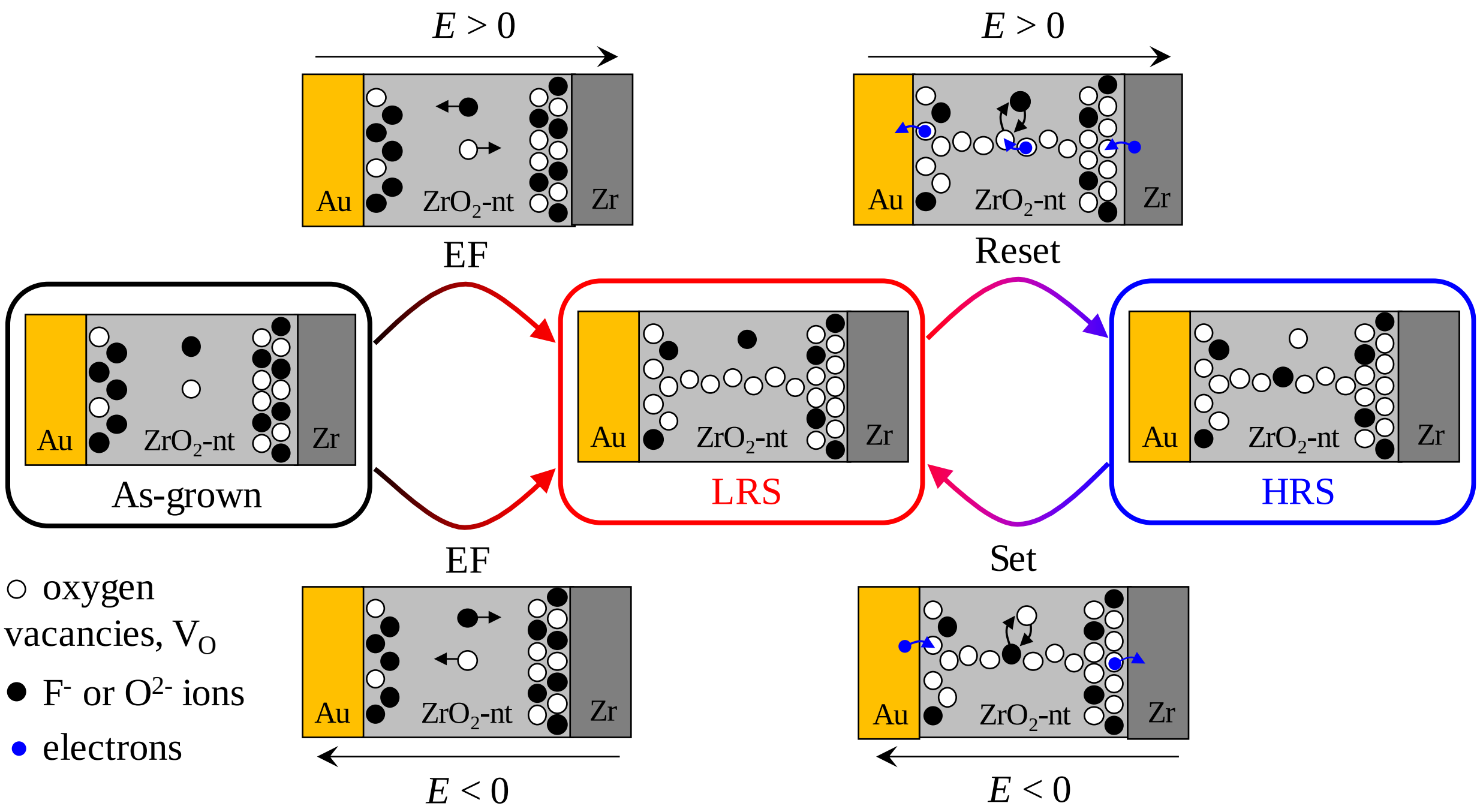

**Figure 8.** Schematic representation of the processes of formation (EF), destruction (Reset), and restoration (Set) of a quantum conductive filament in a $ZrO_2$ nanotube for a bipolar switching mode. Detailed explanations for the figure are given in the text.

## Conclusion

Memristor structures based on a 1.7 μm thick $ZrO_2$ nanotubular layer synthesized by anodic oxidation were fabricated. An analysis of SEM images showed that the oxide layer consists of an ordered array of vertically oriented nanotubes with average internal and external diameters of 55 and 75 nm, respectively.

The *I-V* curves of the Zr/$ZrO_2$-nt/Au memristors investigated exhibit a unipolar and bipolar mechanisms of the *LRS* ↔ *HRS* resistive switching over several tens of complete switching cycles. The ranges of changes in the resistances of the synthesized structures, $R_{HRS} \geq 8$ MΩ and $R_{LRS} \leq 450$ Ω, as well as the ratio $R_{HRS} / R_{LRS} \geq 3 \cdot 10^4$, were determined. Based on the analysis of the *I-V* curves, it can be inferred that the conductivity either of the ohmic type or limited by space charge (SCLC) is realized in the *LRS* and *HRS* states, respectively. The effective carrier mobility $\mu_{eff} = (0.1\text{-}12) \cdot 10^{-3}$ cm$^2$/(V·s) in the studied structure was calculated. The concentration of charge carrier traps based on oxygen vacancies, $N_t = (1\text{-}3) \cdot 10^{15}$ cm$^{-3}$, in the non-stoichiometric nanotubular $ZrO_{2-x}$ layer was estimated. The prospects of using the Zr/$ZrO_2$-nt/Au layered structure as a functional medium for memristor memory elements, as well as possible trends for improving their functional characteristics, are shown.

## Acknowledgements

The work was supported by Minobrnauki research project FEUZ-2020-0059. Authors thank I.N. Bainov for help in XRD measurements.